%% file: delta_q_paper.tex
\documentstyle[multicol,prl,aps,epsfig]{revtex}
\newcommand{\etal}{{\it et al.}}
\newcommand{\np}{Nucl. Phys. }
\newcommand{\prev}{Phys. Rev. }
\newcommand{\nim}{Nucl. Instr. Meth. }

\begin{document}
\title{ Flavor Decomposition of the Polarized Quark Distributions in
  the Nucleon \\ from Inclusive and Semi-inclusive Deep-inelastic
  Scattering}
\author{
\centerline {\it The HERMES Collaboration}\medskip \\
K.~Ackerstaff$^{6}$, 
A.~Airapetian$^{36}$, 
N.~Akopov$^{36}$,
I.~Akushevich$^{7}$,
M.~Amarian$^{26,31,36}$, 
E.C.~Aschenauer$^{7,14,26}$, 
H.~Avakian$^{11}$, 
R.~Avakian$^{36}$, 
A.~Avetissian$^{36}$, 
B.~Bains$^{16}$
% ...
}

\address{
% ...
$^6$DESY, Deutsches Elektronen Synchrotron, 22603 Hamburg, Germany\\
$^7$DESY Zeuthen, 15738 Zeuthen, Germany\\
% ...
$^{11}$Istituto Nazionale di Fisica Nucleare, Laboratori Nazionali di Frascati, 00044 Frascati, Italy\\
% ...
$^{14}$Department of Subatomic and Radiation Physics, University of Gent, 9000 Gent, Belgium\\
% ...
$^{16}$Department of Physics, University of Illinois, Urbana, Illinois 61801, USA\\
% ...
$^{26}$Nationaal Instituut voor Kernfysica en Hoge-Energiefysica (NIKHEF), 1009 DB Amsterdam, The Netherlands\\
% ...
$^{31}$Istituto Nazionale di Fisica Nucleare, Sezione Sanit\`a and Physics Laboratory, Istituto Superiore di Sanit\`a, 00161 Roma, Italy\\
% ...
$^{36}$Yerevan Physics Institute, 375036, Yerevan, Armenia
}
\maketitle
%
\begin{abstract}
  Spin asymmetries of semi-inclusive cross sections for the
  production of positively and negatively charged hadrons have been measured
  in deep-inelastic scattering of polarized positrons on polarized
  hydrogen and $^3$He targets, in the kinematic range $0.023<x<0.6$
  and $1\mbox{ GeV}^2<Q^2<10\mbox{ GeV}^2$.  Polarized quark
  distributions are extracted as a function of $x$ for up $(u+\bar u$)
  and down ($d+\bar d$) flavors. The up quark polarization is positive
  and the down quark polarization is negative in the measured range.
  The polarization of the sea is compatible with zero.  The first
  moments of the polarized quark distributions are presented. The
  isospin non-singlet combination $\Delta q_3$ is consistent with the
  prediction based on the Bjorken sum rule. The moments of the
  polarized quark distributions are compared to predictions based on
  SU(3)$_f$ flavor symmetry and to a prediction from lattice QCD.
\end{abstract}

\begin{multicols}{2}[]
The understanding of the spin structure of the nucleon in terms of
quarks and gluons remains a challenge since it was demonstrated by
EMC~\cite{emc1} and later
experiments~\cite{e142n1,e142n2,e143d,e143p,e154,smc3,smcpre1,smcpre1a,smcpre2,smcpre3,hermesg1n,hermesg1p}
using inclusive deep-inelastic scattering (DIS) that only a 
fraction of the nucleon spin can be attributed to the quark spins and
that the strange quark sea seems to be negatively
polarized~\cite{elkarl}.
These conclusions follow from the extraction of the first moments of
up, down and strange quark spin distributions from the inclusive data
by assuming SU(3)$_f$ flavor symmetry. With semi-inclusive polarized
deep-inelastic scattering experiments, the separate spin contributions
$\Delta q_f$ of quark and antiquark flavors $f$ to the total spin of
the nucleon can be determined as a function of the Bjorken scaling
variable $x$.  Semi-inclusive data can be used to measure the sea
polarization directly and to test SU(3)$_f$ symmetry by comparing the
first moments of the flavor distributions to the SU(3)$_f$
predictions.

Hadron production in DIS is described by the absorption of a virtual
photon by a point-like quark and the subsequent fragmentation into a
hadronic final state. The two processes can be characterized by two
functions: the quark distribution function $q_f(x,Q^2)$,
and the fragmentation function $D_f^h(z,Q^2)$.  The semi-inclusive DIS
cross section $\sigma^h(x,Q^2,z)$ to produce a hadron of type $h$ with
energy fraction $z=E_h/\nu$ is then given by
\begin{equation}
    \label{eq:hadronsig}
    \sigma^h(x,Q^2,z)\propto\sum_f e_f^2 q_f(x,Q^2) D_f^h(z,Q^2).
\end{equation}
The sum is over quark and antiquark types $f=(u,\bar u,d,\bar d,
s,\bar s)$.  In the target rest frame, $E_h$ is the energy of the
hadron, $\nu=E-E'$ and $-Q^2$ are the energy and the squared four-momentum
of the exchanged virtual photon, $E$($E'$) is the energy of the
incoming (scattered) lepton and $e_f$ is the quark charge in units of
the elementary charge.  The Bjorken variable $x$ is calculated from
the kinematics of the scattered lepton according to $x=Q^2/2M\nu$ with
$M$ being the nucleon mass.
It is assumed that the fragmentation process is spin independent,
i.e.~that the probability to produce a hadron of type $h$ from a quark
of flavor $f$ is independent of the relative spin orientations of
quark and nucleon.  The spin asymmetry $A_1^h$ in the semi-inclusive
cross section for production of a hadron of type $h$ by a polarized
virtual photon is given by
\begin{equation}
      A_1^h(x,Q^2,z)={\sum_f e_f^2  \Delta q_f(x,Q^2) D_f^h(z,Q^2)\over
      \sum_f e_f^2 q_f(x,Q^2)D_f^h(z,Q^2)} {(1+R(x,Q^2))\over (1+\gamma^2)}\label{eq:hadronasym}
\end{equation}
where $\Delta q_f(x,Q^2)= q_f^{\uparrow\uparrow}(x,Q^2)-q_f^{\uparrow\downarrow}(x,Q^2)$ is the polarized
quark distribution function and $q_f^{\uparrow\uparrow(\uparrow\downarrow)}(x,Q^2)$ is the
distribution function of quarks with spin orientation parallel
(anti-parallel) to the spin of the nucleon. The ratio
$R=\sigma_L/\sigma_T$ of the longitudinal to transverse photon
absorption cross sections appears in this formula to correct for the
longitudinal component that is included in the experimentally
determined parametrizations of $q_f(x,Q^2)$ but not in $\Delta
q_f(x,Q^2)$.  It is assumed that the ratio of longitudinal to
transverse components is flavor and target independent and that the
contribution from the second spin structure function $g_2(x,Q^2)$ can
be neglected.  The term $\gamma=\sqrt{Q^2}/\nu$ 
is a kinematic factor. Eq.~(\ref{eq:hadronasym}) can be used to extract the
quark polarizations $\Delta q_f(x)/q_f(x)$ from a set of measured
asymmetries on the proton and neutron for positively and negatively
charged hadrons.
    
This paper reports on the extraction of polarized quark distribution
functions from data taken by the HERMES experiment~\cite{hermes1}
using the 27.5 GeV beam of longitudinally polarized positrons
 in the HERA storage ring at DESY, incident on a
longitudinally polarized $^3$He or $^1$H internal gas target.
    
The positron beam at HERA becomes transversely polarized 
by synchrotron radiation emission through its asymmetric
spin-flip probabilities~\cite{sok64}.  The required longitudinal
polarization direction at the HERMES experiment is obtained using spin rotators
located upstream and downstream of the experiment~\cite{bar95}.  The
beam polarization is measured continuously using Compton
backscattering of circularly polarized laser light~\cite{bar93,most}.
The average polarization for the analyzed data was 0.55.  The
fractional statistical error for a single 60~s polarization
measurement was typically 1-2\% and the overall fractional systematic
error was 4.0\% (3.4\%) for the $^3$He (H) measurement, dominated by
the uncertainty in the calibration of the beam polarimeters.
    
The internal target consists of polarized $^3$He (H) gas confined in a
storage cell~\cite{targpaper}, which is a 400~mm long open-ended thin-walled elliptical
tube mounted coaxially with the HERA positron beam.
It was fed by an optically pumped source of polarized $^3$He
atoms~\cite{ref2} in 1995, and by an atomic beam
source of nuclear-polarized hydrogen based on Stern-Gerlach
separation~\cite{abs} in 1996 and 1997.
The tube was constructed of 125 $\mu$m (75
$\mu$m) thick ultra-pure aluminum and
was cooled to typically 25 K (100~K)~\cite{ref2}. This provided a
target with an areal density of approximately $3.3 \times 10^{14}$
$^3$He-atoms/cm$^2$ ($7 \times 10^{13}$ H-atoms/cm$^2$).
During H operation, a drifilm-coated cell~\cite{drifilm} 
was used to minimize wall collision effects. There is
good evidence that recombination is further suppressed by water
deposited on the cell surface during normal operation~\cite{targpaper}.
The
polarization direction was defined by a 3.5~mT (335~mT) magnetic field
parallel to the beam direction and was reversed every 10 (1-2)
minutes.  The polarizations of the $^3$He gas in both the pumping and
the storage cell were measured continuously with optical polarimeters.
The average $^3$He target polarization was 0.46 with a fractional
uncertainty of 5\%.  The relative populations of the hydrogen atomic
states were measured in a Breit-Rabi polarimeter~\cite{brp}. A target
gas analyser was used to measure the atomic and the molecular content
of the hydrogen gas.  The average proton target polarization was 0.86
with a fractional uncertainty of 5\%. The luminosity was measured by
detecting Bhabha-scattered target electrons in coincidence with the
scattered positron.  During the course of a positron fill of typically
8 hours, the current in the ring decreased from typically 40~mA at
injection to about $10$~mA.
    
The HERMES detector is an open-geometry forward spectrometer. A
detailed description is given in Ref.~\cite{specpaper}.  The
geometrical acceptance of $\pm $(40-140)~mrad in the vertical
direction and $\pm 170$~mrad in the horizontal direction allows
detection of hadrons produced in coincidence with the scattered
lepton.  The DIS trigger is formed from a coincidence between
signals in scintillator hodoscope planes and a lead-glass calorimeter.  The
identification of the scattered lepton is accomplished using the
calorimeter, a preshower counter, a transition radiation detector, and
a gas threshold \v{C}erenkov counter. This system provides positron
identification with an average efficiency of 98\% and a hadron
contamination of less than $1\%$.
The threshold \v{C}erenkov counter provides pion identification in a
limited kinematic range.

    
Polarized quark distributions have been extracted from a combination
of inclusive and semi-inclusive asymmetry data on $^3$He and hydrogen.
As the wave function for $^3$He is dominated by the configuration with
the two protons paired to zero spin, most of the asymmetry from $^3$He
is due to the neutron \cite{friar}. The analysis procedure described
in Ref.~\cite{hermesg1n,hermesg1p,thesis,maf} was applied.  The inclusive
(semi-inclusive) asymmetry $A^{(h)}_1$ was extracted from the
measured asymmetry $A_\parallel^{(h)}$ using the relation
\begin{equation}
  \label{eq:a1h}
  A_1^{(h)}={A^{(h)}_\parallel/ \left[ D(1+\gamma\eta) \right]}\ ,
\end{equation} 
where $D$ is the depolarization factor for the virtual photon
and $\eta$ is a kinematic factor as given in
Ref.~\cite{hermesg1p}. In Eq.~(\ref{eq:a1h}) the approximation is used that
the contribution of the second spin structure
function $g_2$ to $A_1^{(h)}$ can be neglected.  In the kinematic
region of our measurement $g_2$ was previously measured to be
consistent with zero for the proton and neutron~\cite{e143g2,e155g2}. In each
kinematic bin the value of $A_{\parallel}^{(h)}$ was extracted from
the measured counting rates using
\begin{equation}
  \label{eq:ahexp}
  A^{(h)}_{\parallel}={N_{(h)}^{\uparrow\downarrow}L^{\uparrow\uparrow} - N_{(h)}^{\uparrow\uparrow}L^{\uparrow\downarrow}
    \over N_{(h)}^{\uparrow\downarrow}L_P^{\uparrow\uparrow} + N_{(h)}^{\uparrow\uparrow}L_P^{\uparrow\downarrow}}\:,
\end{equation}
where $N^{\uparrow\uparrow}$ ($N^{\uparrow\downarrow}$) are the numbers of DIS events for target polarization
parallel (anti-parallel) to the beam polarization, and $N_h^{\uparrow\uparrow}$ ($N_h^{\uparrow\downarrow}$) are
the corresponding numbers of hadrons in coincidence with a DIS event.
Here, $L^{\uparrow\uparrow(\uparrow\downarrow)}$ are the luminosities for each  spin state
corrected for dead time, and $L^{\uparrow\uparrow(\uparrow\downarrow)}_P$ are the luminosities
corrected for dead time and weighted by the product of beam and target
polarizations for each spin state. 
    
After applying data quality criteria and kinematic requirements to select DIS
events ($Q^2> 1$ GeV$^2$, $W^2 > 4$ GeV$^2$
and $y < 0.85$), $2.2 \times 10^6$ ($2.3 \times 10^6$)\\
\end{multicols}\newpage
\begin{figure}[ht]
\begin{center}
\epsfxsize 14 cm {\epsfbox{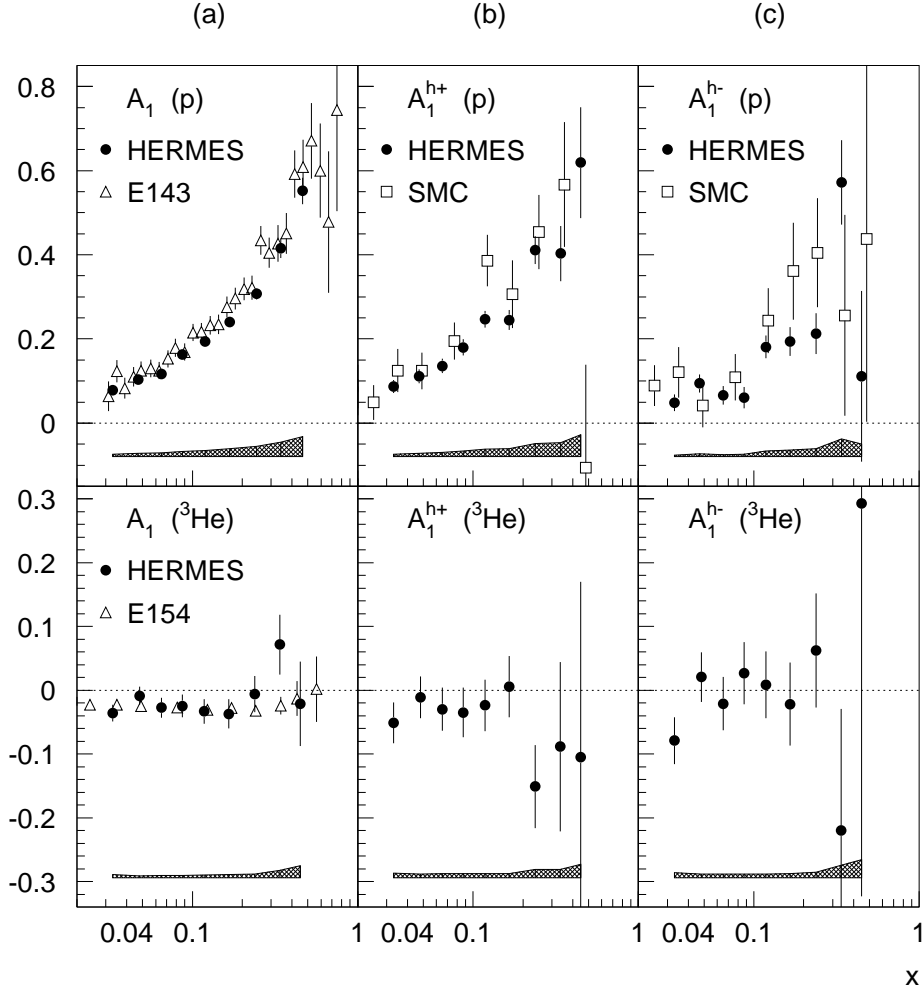}}
\smallskip
\caption{\label{fig:asymmetries}
  The inclusive (a) and semi-inclusive asymmetries for positively (b)
  and negatively (c) charged hadrons on the proton (top) and $^3$He
  (bottom) target.  The inclusive asymmetries are compared to SLAC
  results for $g_1/F_1$ (open triangles).  The hadron asymmetries on
  the proton are compared to SMC results (open squares) truncated to
  the HERMES $x$-range. The data points are given for the measured mean $Q^2$
  at each value of $x$, which is  different for the different experiments.
  The error bars of the HERMES (SLAC and SMC)
  data are statistical (total) uncertainties and the bands are
  systematic uncertainties of the HERMES data.}
\end{center}
\end{figure}
\begin{multicols}{2}[]
\noindent events
were available for analysis on $^3$He (H).  Here, $y=\nu/E$ is the
fractional energy transfer to the
virtual photon and $W$ is the
invariant mass of the initial photon-nucleon system.  The data cover
the ranges $0.023<x<0.6$ and 1~GeV$^2 < Q^2 < 10$~GeV$^2$.  For the
semi-inclusive asymmetries, the hadrons containing information on the
struck quark were distinguished from target region fragments by
requiring each hadron to have a minimum $z$ of $0.2$ and $x_F\approx
2p_L/W$ of $0.1$, where $p_L$ is the longitudinal momentum of the
hadron with respect to the virtual photon in the photon-nucleon center
of mass frame. A minimum $W^2$ cut of 10~GeV$^2$ was additionally
imposed for these events to improve the separation between the current
and target fragmentation regions.  After applying all cuts, $284
\times 10^3$ ($306 \times 10^3$) positive and $178 \times 10^3$ ($175
\times 10^3$) negative hadrons remained for the $^3$He (H) target.
Small corrections were applied to account for charge symmetric
background processes (e.g.  $\gamma\to e^+e^-$).
Smearing corrections were applied to
all data and QED radiative corrections 
were applied only to
the inclusive asymmetries, but not to the semi-inclusive asymmetries,
where the corrections are
negligible~\cite{radcor,radcor2}.

The dominant
sources of systematic uncertainties in the measured asymmetries are:
the target and beam polarization measurements, the uncertainty
assigned for observed yield fluctuations in the $^3$He
data~\cite{hermesg1n} and the systematic uncertainty on $R$~\cite{hermesg1p}. By averaging
over data taken with opposite beam helicities, a possible instrumental 
bias is further reduced.
Fig.~\ref{fig:asymmetries} shows the extracted inclusive asymmetries
and the semi-inclusive asymmetries for positively and negatively
charged hadrons on both targets.
The measured spin asymmetries $A_1^h(x,Q^2,z)$ were integrated in each
$x$ bin over the corresponding $Q^2$-range and the $z$-range from 0.2
to 1 to yield $A_1^h(x)$.
Also shown are inclusive results
measured at a similar energy at SLAC~\cite{e143p,e154,e143g2,yuri}
and 
hadron asymmetries on hydrogen measured by SMC~\cite{smcsemi}.  The
data are in agreement within the quoted uncertainties.
The agreement of the HERMES data with the
SMC data, taken at 6-12 times higher average $Q^2$, shows that
the semi-inclusive asymmetries
are $Q^2$ independent within the present accuracy of the experiments.

Eq.~(\ref{eq:hadronasym}) is used to extract polarized quark
distribution functions from semi-inclusive asymmetries.
It can be written as
\begin{equation}
  \label{eq:hadronpuri}
  A_1^{h}(x)=\sum_f P_f^{h}(x){\Delta q_f(x)\over
    q_f(x)} {(1+R(x))\over (1+\gamma^2)}
\end{equation}
where $P_f^h(x)$ are the integrated purities~\cite{maf,bruins} defined
as
\begin{equation}
  \label{eq:purity}
  P_f^h(x)={ e_f^2  q_f(x) \int_{0.2}^1 D_f^h(z)\, dz\over
    \sum_{f'} e_{f'}^2 q_{f'}(x)\int_{0.2}^1 D_{f'}^h(z')\, dz'}.
\end{equation}
The inclusive asymmetry $A_1$ is similarly expressed by replacing
$P_f^h$ by $ P_f$ where $P_f(x)= e_f^2 q_f(x) / \sum_{f'} e_{f'}^2
q_{f'}(x) $.  After integrating over $z$,
Eq.~(\ref{eq:hadronpuri}) together with the
corresponding inclusive case can be written in matrix form
\begin{equation}
  \vec{A}(x) = {\cal P}(x) \cdot \vec{Q}(x)
\label{eq:matrixeqn}
\end{equation}
where the vector
$\vec{A}=(A_{1p},A_{1p}^{h^+},A_{1p}^{h^-},A_{1He},A_{1He}^{h^+},A_{1He}^{h^-})$
contains as elements the measured asymmetries. The vector $\vec{Q}(x)$
contains the quark and antiquark polarizations.  The matrix $\cal P$
contains the effective integrated purities for the proton and $^3$He
as well as the $(1+R)/(1+\gamma^2)$ factor.  These purities describe the probability
that the virtual photon hit a quark of flavor $f$ when
a hadron of type $h$ is detected in the experiment.
They include the effects of the acceptance
of the experiment and have been determined with a Monte Carlo
simulation using the LUND string fragmentation model~\cite{lund}, a
model of the detector, the CTEQ Low--$Q^2$
parametrizations~\cite{LAI97x} for the unpolarized parton
distributions and values for $R$ from Ref.~\cite{rref}. The LUND
fragmentation parameters were tuned to fit the measured hadron
multiplicities.  For the $^3$He data, a correction was applied for the
non-zero polarization of the protons of $-0.028\pm 0.004$ and the
neutron polarization of $0.86\pm 0.02$~\cite{nuclcorr}.
Eq.~(\ref{eq:matrixeqn}) can be solved for $\vec{Q}(x)$ by minimizing
\begin{equation}
  \chi^2 = \left( \vec{A} - {\cal P} \cdot \vec{Q} 
  \right)^{\mathrm{T}} {\cal V}_{\vec A}^{-1}
  \left( \vec{A} - {\cal P} \cdot \vec{Q} \right).
  \label{eq:chi2eqn}
\end{equation}
where ${\cal V}_{\vec A}$ is the covariance matrix of the
asymmetry vector $\vec A$.  In the fit procedure constraints were
imposed on the sea polarization to improve statistical significance.
In view of rather ambiguous theoretical model
predictions~\cite{diakonv,fries}, two alternatives were chosen for
relating the spin distributions of the sea flavors.
As a first possibility it was assumed that the polarization
${\Delta q_{s}(x)/ q_{s}(x)}$ of sea quarks is independent of
flavor:
\begin{equation}
  \label{eq:seacond}
  {\Delta u_{s}(x)\over u_{s}(x)}
  =      {\Delta d_{s}(x)\over d_{s}(x)}
  =      {\Delta  s(x)\over s(x)}
  =      {\Delta \bar u(x)\over \bar u(x)}
  =      {\Delta \bar d(x)\over \bar d(x)}
  =      {\Delta \bar s(x)\over \bar s(x)}.
\end{equation}
This approach is used for all calculations unless otherwise stated.
As a second approach, a pure singlet spin distribution of the sea is
considered:
\begin{equation}
  \label{eq:seacond2}
  {\Delta u_{s}(x)}
  =      {\Delta d_{s}(x)}
  =      {\Delta  s(x)}
  =      {\Delta \bar u(x)}
  =      {\Delta \bar d(x)}
  =      {\Delta \bar s(x)}.
\end{equation}

\begin{figure}[t]
\begin{center}
\epsfxsize 8 cm {\epsfbox{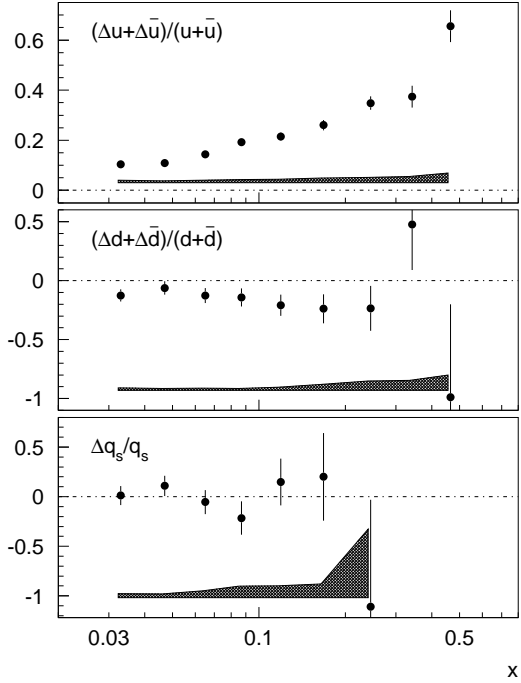}}
\smallskip
\begin{minipage}[r]{\linewidth}
\caption{\label{fig:polarisations}
  The flavor decomposition $(\Delta u(x)+\Delta \bar{u}(x))/$ $(u(x)+\bar{u}(x))$, 
  $(\Delta d(x)+\Delta \bar{d}(x))/(d(x)+\bar{d}(x))$ 
  and ${\Delta q_{s}(x)}/{q_{s}(x)}$ of the quark polarization as a function
  of $x$, derived from the HERMES inclusive and semi-inclusive
  asymmetries. The sea polarization is assumed to be flavor symmetric
  in this analysis. The error bars shown are the statistical and the bands
  represent the systematic uncertainties.}
\end{minipage}
\end{center}
\end{figure}

The flavor decomposition is obtained by solving
Eq.~(\ref{eq:matrixeqn}) for a vector $\vec Q$, which contains the sum
of quarks and antiquarks
\begin{equation}
  \label{eq:flavsep}
  \vec{Q}=\left(
    \frac{\Delta u (x)+ \Delta \bar u(x)}{u (x)+ \bar u(x)},
    \frac{\Delta d (x)+ \Delta \bar d(x)}{d (x)+ \bar d(x)}, 
    \frac{\Delta s (x)+\Delta \bar s(x)}{s (x)+ \bar s(x)}\right) ,
\end{equation}
where due to assumption (\ref{eq:seacond}) the polarizations of the
strange quarks and of the total sea are equal: $(\Delta s(x) +\Delta
\bar s(x))/(s (x)+ \bar s(x))=\Delta q_{s}(x)/ q_{s}(x)$. For $x>0.3$
the sea polarization is set to zero and the corresponding effect on
the results for the non-sea polarizations is included in their
systematic uncertainties. Fig.~\ref{fig:polarisations} shows the
results.  The up quark polarizations are positive and the down quark
polarizations are negative over the measured range of $x$. Their
absolute values are largest at large $x$ and remain different from
zero in the sea region.  The sea polarization is compatible with zero
over the measured range of $x$. The overall $\chi^2$ per degree of freedom
of the fit is 1.1.

The systematic uncertainties, shown by the shaded band in
Fig.~\ref{fig:polarisations}, were determined from the uncertainties
on the measured asymmetries, the unpolarized parton distributions
and the purities.  The uncertainty on the
unpolarized parton distributions was derived by comparing different
parametrizations~\cite{LAI97x,GLU96x} of the world data.  The
uncertainty coming from the symmetry assumption of the sea
polarization was derived by comparing the results produced by
Eqs.~(\ref{eq:seacond}) and (\ref{eq:seacond2}) respectively.
The fitted quark polarizations change by
typically less than $0.01$ when the assumption in Eq.~(\ref{eq:seacond})
is replaced by (\ref{eq:seacond2}).  The uncertainty in the purities
was determined by comparing different fragmentation
models~\cite{fragm1,fragm2} and varying the fragmentation parameters
in the Monte Carlo code.

\begin{figure}[th]
\begin{center}
\epsfxsize 8 cm {\epsfbox{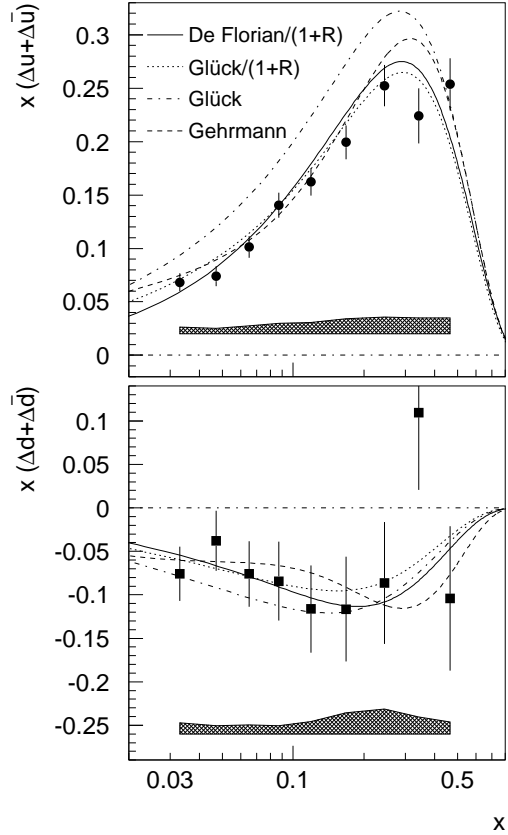}}
\smallskip
\begin{minipage}[r]{\linewidth}
\caption{\label{fig:spindistr}
  The quark spin distributions at $Q^2=2.5$~GeV$^2$ separately for
  $x(\Delta u(x)+\Delta \bar{u}(x))$ and $x(\Delta d(x)+\Delta \bar{d}(x))$ as a
  function of $x$. They are compared to different sets of
  parametrizations which correspond to the following publications:
  De Florian {\it et al.}~($0.1<\Delta G<0.8$, LO)~\protect\cite{Flori}, {Gehrmann and
    Stirling} ('Gluon A', LO)~\protect\cite{GER96}, and Gl\"uck {\it et al.}~('Standard
  Scenario', LO)~\protect\cite{GLU96}.  The {De Florian} and {Gl\"uck}
  parametrizations are corrected by a factor $(1+R)$ to allow for a
  direct comparison. The error bars shown are the statistical and the
  bands represent the systematic uncertainties.}
\end{minipage}
\end{center}
\end{figure}

The polarized quark distributions $\Delta q_f(x)$ were determined by
forming the products of the polarizations $\Delta q_f(x)/q_f(x)$ and
the unpolarized parton distributions from Ref.~\cite{LAI97x} at
$Q^2=2.5$~GeV$^2$. It was assumed that the polarization is independent
of $Q^2$ within the $Q^2$ range of this measurement. This assumption is
justified by the weak $Q^2$ dependence predicted by QCD and by the
experimental result that there is no significant $Q^2$ dependence
observed in the inclusive asymmetries~(see e.g.
Ref.~~\cite{hermesg1p}) and in the semi-inclusive asymmetries as shown
in Fig.~\ref{fig:asymmetries}.  The results for the up and down
distributions are shown in Fig.~\ref{fig:spindistr} and compared with
different sets of parametrizations of world data in leading order
QCD~\cite{Flori,GER96,GLU96}. Parametrizations that were fitted to
spin asymmetries $A_1$ under the assumption $R=0$ do not fit the
HERMES data for $x(\Delta u(x)+\Delta \bar{u}(x))$.
They can be brought into agreement with the HERMES results
by dividing by $1+R$.
Fig.~\ref{fig:spindistr} demonstrates the size of the effect for the
parametrization by Gl\"uck {\it et al.}. Parametrizations that are derived
from fits to $g_1$ instead of $A_1$ (e.g.~{Gehrmann and
  Stirling}) do not need this
correction.

\begin{figure}[ht]
\begin{center}
\epsfxsize 8 cm {\epsfbox{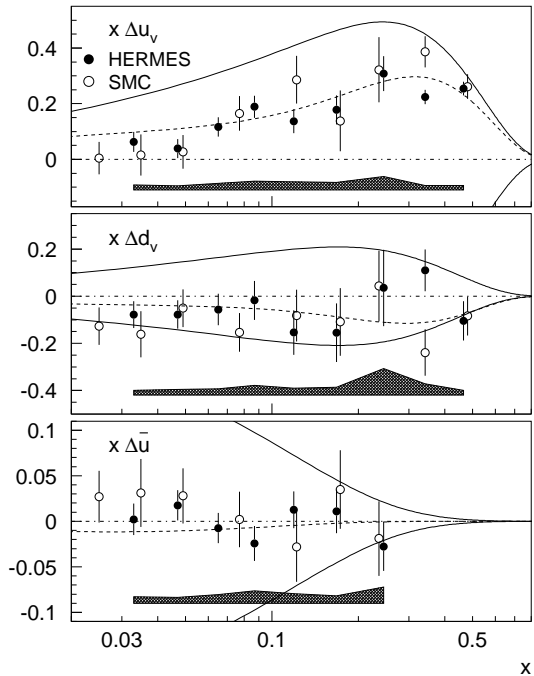}}
\smallskip
\begin{minipage}[r]{\linewidth}
\caption{\label{fig:spindistrv}
  The spin distributions at $Q^2=2.5$~GeV$^2$ separately for the
  valence quarks ${x\Delta u_v (x)}$, ${x\Delta d_v(x)}$ and the sea quarks
  $ {x\Delta \bar u(x)}$ as a function of $x$.  The error bars shown are the
  statistical and the bands the systematic uncertainties.  The
  distributions are compared to results from SMC, ex\-tra\-po\-la\-ted to
  $Q^2=2.5$~GeV$^2$. The error bars of the SMC result correspond to
  its total uncertainty. The solid lines indicate the positivity limit
  and the dashed lines are the parametrization from {Gehrmann and
  Stirling} ('Gluon A', LO)~\protect\cite{GER96}.}
\end{minipage}
\end{center}
\end{figure}

The upper plots in Fig.~\ref{fig:spindistrv} show the polarized
valence quark distributions $x\Delta u_v(x)$ and $x\Delta d_v(x)$,
derived from the relation $\Delta q_v(x)=(\Delta q(x)+\Delta \bar
q(x))-2 \Delta \bar q(x)$. Since\\
\noindent  for scattering off sea quarks the contribution from $u_s$ and $\bar u$ quarks
dominates, the polarized $x\Delta \bar u(x)$
sea distribution is shown in the lower plot.
Fig.~\ref{fig:spindistrv} includes results from SMC~\cite{smcsemi}
obtained at $Q^2=10$~GeV$^2$, which are shown here for the $x$-range
explored by HERMES and which are extrapolated to $Q^2=2.5$~GeV$^2$ by
assuming a $Q^2$-independent polarization $\Delta q(x)/q(x)$. The SMC
results are derived under the assumption presented in
Eq.~(\ref{eq:seacond2}) rather than (\ref{eq:seacond}).  The
positivity limit and a parametrization of data from Ref.~\cite{GER96}
are included in Fig.~\ref{fig:spindistrv}.  The parametrization and
the SMC results are consistent with the HERMES results within the
statistical and systematic uncertainties.
The uncertainties of the\\
\noindent HERMES data for $x\Delta u_v(x)$ and $x\Delta
\bar u(x)$  are much smaller than for the SMC data.

\begin{figure}[ht]
\begin{center}
\epsfxsize 7.9 cm {\epsfbox{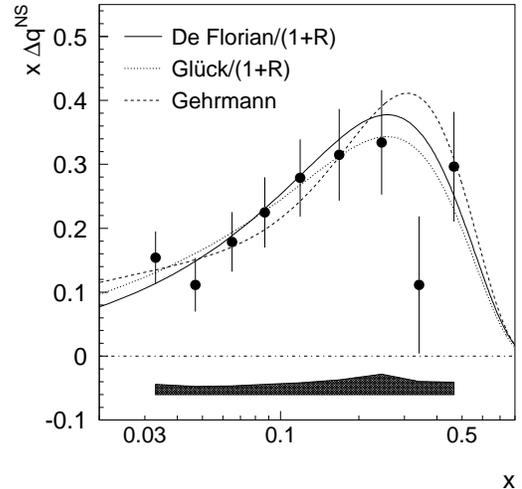}}
\smallskip
\begin{minipage}[r]{\linewidth}
\caption{The  non-singlet contribution $x\Delta q^{NS}(x)$ at $Q^2=2.5$~GeV$^2$.
  The result is compared to the same sets of parametrizations as in
  Figure~\ref{fig:spindistr}. The error bars are the statistical and the band
  the systematic uncertainties.
\label{fig:nsinc}}
\end{minipage}
\end{center}
\end{figure}

In the quark parton model the isospin non-singlet combination $\Delta
q^{NS}(x)=\Delta u(x)+\Delta \bar u(x)-\Delta d(x) -\Delta \bar d (x)$
is directly related to the spin structure functions according to
$\Delta q^{NS}(x)= 6( g_1^p(x)-g_1^n(x))$.  Fig.~\ref{fig:nsinc}
illustrates that the HERMES result for  $\Delta q^{NS}(x)$  is
in good agreement with
parametrizations of other published inclusive data.

The first and second moments of spin distributions have been
calculated and compared to other experimental data and to model
predictions.  In the measured $x$-region, the integral $\Delta q_f$ is
obtained as
\begin{equation}
  \label{eq:measint}
\Delta q_f = \sum_i \left( {\Delta q_f\over{q_f}}\bigg|_i \,\int_{x_i}^{x_{i+1}}\!
q_f(x)\> dx \right),
\end{equation}
where $(\Delta q_f/q_f)|_i$ is constant within each bin
$(x_i,x_{i+1})$ and $q_f(x)$ is a parametrization given in
Ref.~\cite{LAI97x}.  Outside the measured region $0.023<x<0.6$,
extrapolations are required.  There is no clear prediction for the low-$x$
\\
\noindent extrapolation~\cite{e154}.  For comparison with previous
measurements we quote the integrals assuming a simple Regge
parametrization\cite{reggeexp,regge} of $\Delta q_f (x)\propto
x^{-\alpha}$ with $\alpha=0$ fitted to the data for $x < 0.075$.
Due to the strong model dependence of the result, no error is quoted
for the extrapolation.  The extrapolation to $x = 1$ was obtained by
fitting the functional form from Ref.~\cite{GLU96} to our data. The
fit is constrained to fulfill the positivity limit $|\Delta
q_f(x)/q_f(x)|\leq (1+\gamma^2)/(1+R(x))$. The systematic uncertainty of the
high-$x$ extrapolation is small and included in the quoted uncertainty
of the total integral.

The results are listed in Table~\ref{tab:integraltheo}.  The spin
carried by up, down and strange quarks is compared to results from the
SU(3)$_f$ analysis of the inclusive data. The comparison
is performed by using the $\Delta q_0$ value and the QCD corrections
from Ref.~\cite{elkarl}.  We obtain values for the up, down and
strange contributions which differ 
from our  semi-inclusive result.
The sea quark contribution is found to be close to zero in this
semi-inclusive analysis whereas the strange quark sea is significantly
negative in the inclusive analysis. Neither result represents a direct
measurement of the strange spin distribution but rather depends on
the assumptions of SU(3)$_f$ symmetry for the inclusive case and on
the sea symmetry condition (Eq.~(\ref{eq:seacond})) for the
semi-inclusive case. To interpret the differences that are observed in
the contributions from up and down quarks, the flavor distributions
have been separated into SU(3)$_f$ singlet $(\Delta q_0)$, triplet
$(\Delta q_3)$, and octet $(\Delta q_8)$ contributions, as shown in
Table~\ref{tab:integraltheo}.  The  total spin integral \\
\end{multicols}\newpage
\begin{table}[ht]
  \caption{
    The integrals of various spin distributions. The results are given
    for the measured  region $0.023<x<0.6$, for the low-$x$
    extrapolation and for the total integral.
    Note that the entry for $\Delta s +\Delta \bar{s}$ does not represent a
    direct measurement of the strange sea but relies on the assumption in
    Eq.~(\ref{eq:seacond}) (see text).
    An uncertainty of the Regge-type extrapolation at low-$x$
     is not included in the quoted error
    for the total integral.  The items $ x \Delta
    u_{v}$ and $ x \Delta d_{v}$ denote the second moments. $\Delta
    q_8^*$ uses Eq.~(\ref{eq:seacond2}) whereas all other quantities
    use Eq.~(\ref{eq:seacond}) as symmetry condition.
    The HERMES results 
    are given for $Q^2=2.5$~GeV$^2$.  The $Q^2$ values of
    the predictions  are quoted in
    GeV$^2$.    } 
   \label{tab:integraltheo}
\begin{center}
  \begin{minipage}[r]{14cm}
  \begin{tabular}{|c||r|r|r|rl|c|}
    & \multicolumn{1}{c|}{measured region}
    & \multicolumn{1}{c|}{low-$x$}
    & \multicolumn{1}{c|}{total integral}
    &\multicolumn{2}{c|}{prediction}
    &Q$^2$\\                          
    \hline\hline
    $\Delta u +\Delta \bar{u}$ &
   $  0.51\pm   0.02\pm   0.03$ &$0.04 $
  & $ 0.57\pm 0.02\pm 0.03 $ &
    $ 0.66\pm 0.03 $&SU(3)              &2.5\\
    $\Delta d +\Delta \bar{d}$ &
    $-0.22\pm 0.06\pm 0.05 $ &  $-0.03  $ &
    $-0.25\pm 0.06 \pm 0.05$ & 
    $ -0.35\pm 0.03 $&SU(3) &2.5 \\
    $\Delta s +\Delta \bar{s}$ &
    $-0.01\pm     0.03\pm 0.04 $ & $ 0.00$ &
    $-0.01\pm     0.03\pm 0.04 $ &
    $-0.08\pm     0.02 $&SU(3) &2.5 \\
    \hline
    $\Delta \bar{u}$ &
    $ -0.01 \pm 0.02 \pm 0.03      $ & $0.00$ &
    $ -0.01\pm 0.02\pm 0.03   $ &&&\\
    $\Delta \bar{d}$ &
    $ -0.02\pm 0.03 \pm 0.04  $ & $0.00 $&
    $ -0.01\pm 0.03 \pm 0.04                       $&&&\\
    \hline
    $\Delta q_0$ &
    $ 0.28\pm 0.04\pm 0.09 $    &$0.01 $&
    $  0.30\pm 0.04\pm 0.09                     $   &
    $ 0.23\pm 0.04 $&SU(3)
    \cite{elkarl} &2.5\\
    $\Delta q_3$ &
    $ 0.74\pm     0.07\pm 0.06 $ &$0.07$ &
    $ 0.84\pm     0.07\pm 0.06                           $  &
    $1.01 \pm     0.05 $
    & Bjorken&2.5\\
    $\Delta q_8$ &
    $ 0.32\pm     0.09\pm 0.10 $ & $  0.01 $ &
    $  0.32\pm     0.09\pm 0.10                          $  &
    $0.35 \pm  0.07$ &non-SU(3)\cite{licht}&2.5\\
    $\Delta q_8^*$ &
    $0.33\pm 0.10\pm 0.11     $ & $  0.00 $ &
    $0.33\pm 0.10\pm 0.11     $  &
    $0.46 \pm  0.03$ &$F\& D$&2.5\\
    \hline
    $\Delta u_{v}$ &
    $ 0.52\pm 0.05\pm 0.08 $ & $0.03$ &
    $ 0.57\pm 0.05\pm 0.08                      $ &
 $0.84 \pm 0.05  $   &Lattice \cite{goeck} &5\\
    $\Delta d_{v}$ &
    $-0.19\pm 0.11\pm 0.13 $ & $ -0.03  $  &
    $-0.22\pm 0.11\pm 0.13                      $&
 $-0.25 \pm 0.02  $ &Lattice \cite{goeck}  &5\\
    \hline
    $ x\Delta u_{v}$ &
    $ 0.12 \pm 0.01 \pm 0.01  $ &$0.00 $  &
    $ 0.13 \pm 0.01 \pm 0.01  $& $ 0.198 \pm 0.008 $ &Lattice \cite{goeck}  &4 \\
     $ x \Delta d_{v}$ &
    $-0.02 \pm 0.02 \pm 0.02  $ &$ 0.00  $&
    $-0.02 \pm 0.02 \pm 0.02  $ & $-0.048  \pm 0.003  $&Lattice \cite{goeck} &4
  \end{tabular}
\end{minipage}
\end{center}
\end{table}
\begin{multicols}{2}[]
\noindent $\Delta q_0=\int_0^1
( \Delta u(x)+\Delta \bar u(x)+\Delta d(x)+\Delta \bar d(x)+\Delta
s(x)+\Delta \bar s(x))\, dx=  0.23\pm 0.04 $
agrees well with the HERMES result $0.30\pm 0.04(stat.)\pm 0.09(syst.) $.
The large theoretical uncertainties on the extrapolation at low $x$
are not included here.  The triplet contribution $\Delta q_3=\int_0^1
\Delta q^{NS}(x)\, dx=0.84\pm 0.07(stat.)\pm 0.06(syst.)$ is directly
related to the Bjorken sum rule~\cite{bjsr} according to
\begin{eqnarray}
 \int_0^1\Delta q^{NS}(x)\, dx 
  &=& \left|{g_a\over g_v}\right|\times C_{\mbox{\footnotesize QCD}}
\end{eqnarray}
with the  QCD correction $C_{\mbox{\footnotesize QCD}}$ according to
Ref.~\cite{elkarl,larin,clorob}. 
Higher twist corrections are expected to be small~\cite{HT,HT1,HT2,HT3} and have
been neglected.  The semi-inclusive result agrees with the prediction
of the Bjorken sum rule of $1.01 \pm 0.05$ obtained including an estimate of
the QCD correction in 4$^{th}$ order in $\alpha_s=0.35 \pm  0.04$ for
$Q^2=2.5$~GeV$^2$. The large theoretical error of the Bjorken sum rule
prediction comes from the uncertainties in $\alpha_s$.

The Ellis-Jaffe sum rule~\cite{ejsr}, which is based on SU(3)$_f$
flavor symmetry and on the assumption of a zero polarization of
strange quarks, has been found to be
violated~\cite{emc1,e142n1,e142n2,e143d,e143p,e154,smc3,smcpre1,smcpre1a,smcpre2,smcpre3,hermesg1n,hermesg1p}.
Models to explain this discrepancy invoke either SU(3)$_f$
symmetry breaking, a large negative strange quark polarization, or
SU(3)$_f$-asymmetric polarized sea distributions. Semi-inclusive data
provide a test of such models. This is illustrated for two
examples: (i) a model with symmetric sea (Eq.~(\ref{eq:seacond2})) and
unbroken SU(3)$_f$ and (ii) a model which is not SU(3)$_f$ symmetric
combined with a flavor asymmetric sea according to
Eq.~(\ref{eq:seacond}).

In the first case the octet combination $\Delta q_8 =\int_0^1 (\Delta u
(x)+ \Delta \bar u(x)+\Delta d (x)+ \Delta \bar d(x) -2(\Delta s(x)
+\Delta \bar s(x)))\, dx $ can be related to the hyperon decay
constants $F$ and $D$ according to $\Delta q_8= (3F-D)\times
C_{\mbox{\footnotesize QCD}}=0.46 \pm 0.03$, where
$C_{\mbox{\footnotesize QCD}}$ is taken from Ref.~\cite{clorob}. The
semi-inclusive result yields $\Delta q_8=0.33\pm 0.10(stat.)\pm
0.11(syst.) $ which is lower than the prediction, but still 
consistent.

\begin{figure}[ht]
\begin{center}
\epsfxsize 8 cm {\epsfbox{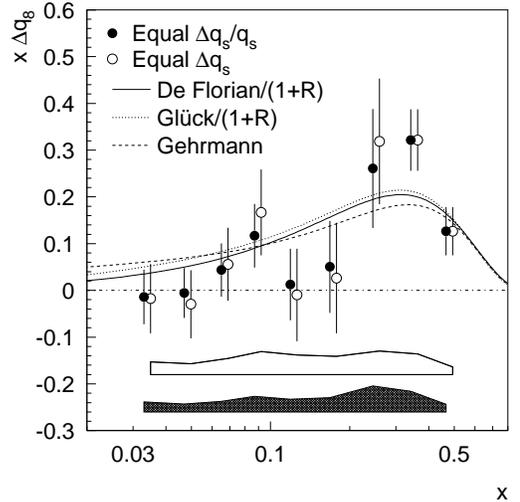}}
\smallskip
\begin{minipage}[r]{\linewidth}
\caption{\label{fig:a8}
  The SU(3)$_f$ octet combination $x\Delta q_8(x)$ at $Q^2=2.5$~GeV$^2$
  assuming a flavor symmetric sea (open circles) or a flavor
  independent polarization (full circles).  The data are compared to
  the same sets of parametrizations as in Figure~\ref{fig:spindistr}.
  The error bars shown are the statistical and the upper (lower) band
  shows the systematic uncertainties which correspond to the open (full)
  circles.}
\end{minipage}
\end{center}
\end{figure}

\end{multicols}\newpage

{\begin{table}[ht]
\begin{minipage}[r]{\linewidth}
  \caption{Comparison of the HERMES integrals of valence and sea  spin
    distributions with SMC results. The SMC values
    are  extrapolated to $Q^2=2.5$~GeV$^2$
    and integrated over the HERMES $x$-range of  $0.023<x<0.6$. 
   } 
   \label{tab:integralmeas}
\begin{center}
  \begin{minipage}[r]{7cm}
  \begin{tabular}{|c||r|r|}
    & \multicolumn{1}{c|}{HERMES}
    & \multicolumn{1}{c|}{SMC}\\
    \hline\hline
    $\Delta \bar{u}$ &
   $ -0.01 \pm 0.02 \pm 0.03      $  & $0.02\pm 0.03\pm     0.02$\\
    $\Delta \bar{d}$ &
     $ -0.02\pm 0.03 \pm 0.04  $ & $0.02\pm 0.03\pm 0.02$\\
    \hline
    $\Delta u_{v}$ &
   $ 0.52\pm 0.05\pm 0.08 $ & $0.59\pm 0.08\pm 0.07$\\
    $\Delta d_{v}$ &
   $-0.19\pm 0.11\pm 0.13 $ & $-0.33\pm 0.11\pm 0.09$\\
    \hline
    $ x\Delta u_{v}$ &
    $ 0.12 \pm 0.01 \pm 0.01  $ & $0.15 \pm  0.02 \pm 0.01 $\\
    $ x \Delta d_{v}$ &
   $-0.02 \pm 0.02 \pm 0.02  $ &    $-0.05 \pm 0.02 \pm 0.02 $
  \end{tabular}
\end{minipage}
\end{center}
\end{minipage}
\end{table}}
\begin{multicols}{2}[]
\noindent   A similar deviation shows up at low $x$ in the
comparison of the semi-inclusive result of $\Delta q_8(x)$ with the
fits from Ref.~\cite{Flori,GER96,GLU96} which are dominated by
inclusive data (see Fig.~\ref{fig:a8}).

In the second case we assume a flavor asymmetric polarized sea 
according to Eq.~(\ref{eq:seacond}) and obtain from the semi-inclusive
analysis the result $\Delta
q_8=0.32\pm 0.09(stat.)\pm 0.10(syst.) $. Following Ref.~\cite{licht}
$\Delta q_8 $ can be calculated according to
\begin{eqnarray}
  \label{eq:lichteq}
\Delta q_8&=&{2\epsilon\Delta q_0+3(3F-D)
 C_{\mbox{\footnotesize
    QCD}}\over 3+2\epsilon }
\end{eqnarray}
where $\epsilon$ is estimated from the averaged ratio of distribution
functions in Ref.~\cite{LAI97x} according to $ 1+\epsilon= \left<{\Delta
  \bar d(x)/ \Delta \bar s(x)}\right>= \left<{\bar d(x)/ \bar s(x)}\right>= 4.8\pm 0.4$.
The model yields $\Delta q_8=0.35\pm 0.07 $.

The results are compatible with both  assumptions and do not allow for
a definite statement about the question whether  the polarized
quark distributions violate SU(3) flavor symmetry.  A direct measurement of
the strange sea is required for a final conclusion about the reason
for the violation of the Ellis-Jaffe sum rule.

The first and second moments of the valence spin distributions $\Delta
u_v(x)$ and $\Delta d_v(x)$ have been extracted (see
Table~\ref{tab:integraltheo}) and are compared to SMC
results~\cite{smcsemi} (see Table~\ref{tab:integralmeas}). For this
comparison the SMC values have been extrapolated to $Q^2=2.5$~GeV$^2$
and integrated according to Eq.~(\ref{eq:measint}) in the HERMES
$x$-range.  There is good agreement with the SMC data.  Significant
deviations are observed between the measured first and second moments
of $\Delta u_v(x)$ from predictions by lattice QCD~\cite{goeck}.
Notice however, that the QCD calculation has been performed in the
quenched approximation.
The relative factor between the measured value and the lattice
value for the second moment of $\Delta u_v(x)$ is similar to that observed
in the unpolarized case~\cite{goe2}.

In summary, inclusive and semi-inclusive spin asymmetries on
longitudinally polarized hydrogen and $^3$He targets were measured and
used to extract the individual quark spin polarizations for up and
down quarks, and for valence and sea quarks.  The up distributions are
positive, the down distributions are negative and the sea quarks show
no significant polarization.  The values for the integrals of the
polarized quark distributions have been determined at
$Q^2=2.5$~GeV$^2$. The results for the flavor-separated first moments
differ  from the inclusive results from Ref.~\cite{elkarl}
that are based on
SU(3)$_f$ flavor symmetry. The measured value of the octet combination $\Delta
q_8(x)$ is lower but still consistent with the value  $3F-D$ predicted by SU(3)$_f$
symmetry arguments.  It provides a test for models trying to explain
the violation of the Ellis-Jaffe sum rule.  The first moment of the
non-singlet combination of the polarized quark distributions agrees
with predictions from the Bjorken sum rule. Predictions from a
quenched lattice QCD calculation overestimate the first and second
moments of the polarized valence up quark distribution.

We gratefully acknowledge the DESY management for its support and
the DESY staff and the staffs of the collaborating institutions.
This work was supported by
the FWO-Flanders, Belgium;
the Natural Sciences and Engineering Research Council of Canada;
the INTAS, HCM, and TMR network contributions from the European Community;
the German Bundesministerium f\"ur Bildung, Wissenschaft, Forschung
und Technologie; the Deutscher Akademischer Austauschdienst (DAAD);
the Italian Istituto Nazionale di Fisica Nucleare (INFN);
Monbusho, JSPS, and Toray
Science Foundation of Japan;
the Dutch Foundation for Fundamenteel Onderzoek der Materie (FOM);
the U.K. Particle Physics and Astronomy Research Council; and
the U.S. Department of Energy and National Science Foundation.



\end{multicols} 
\clearpage

%
%

%
%

\begin{table}
\centering
\caption{Bin boundaries in the variable $x$.}
  \bigskip
  \begin{tabular}{|c||c|c|c|c|c|c|c|c|c|}
  Bin number & 1  & 2  & 3 & 4 & 5 & 6 & 7 & 8 & 9 \\
  \hline
  $x$-range  
  & 0.023-0.040 
  & 0.040-0.055
  & 0.055-0.075
  & 0.075-0.10
  & 0.10-0.14
  & 0.14-0.20
  & 0.20-0.30
  & 0.30-0.40
  & 0.40-0.60\\
  \end{tabular}
\end{table}

%
%

\begin{table}
\centering
\caption{The values of the inclusive proton asymmetry $A_1$ and 
semi-inclusive proton asymmetries for positively charged hadrons 
($A_1^{h^+}$) and negatively charged hadrons ($A_1^{h^-}$). The values 
are quoted at the average measured values of $x$ and $Q^2$ in each $x$-bin. 
The first error is statistical, the second one systematic.  
  \label{tab:a1_proton}}
  \bigskip
  \begin{tabular}{|c|c|r|}
  $\langle x \rangle$ & 
  $\langle Q^2\rangle\quad \left[({\rm GeV}/c)^2\right]$ & 
  $A_1 \pm$ stat. $\pm$ syst. \\
  \hline
  \input{a1.proton.tex}
  \end{tabular}

  \bigskip
  \begin{tabular}{|c|c|r|} 
  $\langle x \rangle$ & 
  $\langle Q^2\rangle\quad \left[({\rm GeV}/c)^2\right]$ & 
  $A_1^{h^+} \pm$ stat.$\pm$ syst. \\
  \hline
  \input{a1h+.proton.tex}
  \end{tabular}

  \bigskip
  \begin{tabular}{|c|c|r|} 
  $\langle x \rangle$ & 
  $\langle Q^2\rangle\quad \left[({\rm GeV}/c)^2\right]$ & 
  $A_1^{h^-} \pm$ stat.$\pm$ syst. \\
  \hline
  \input{a1h-.proton.tex}
  \end{tabular}
\end{table}

%
%

\begin{table}
  \centering
  \caption{The values of the inclusive $^3$He asymmetry $A_1$ and 
  semi-inclusive $^3$He asymmetries for positively charged hadrons 
  ($A_1^{h^+}$) and negatively charged hadrons ($A_1^{h^-}$). The values 
  are quoted at the average measured values of $x$ and $Q^2$ in each 
  $x$-bin. The first error is statistical, 
  the second one systematic.  
  \label{tab:a1_helium}}
  \bigskip
  \begin{tabular}{|c|c|r|}
  $\langle x \rangle$ & 
  $\langle Q^2\rangle\quad \left[({\rm GeV}/c)^2\right]$ & 
  $A_1 \pm$ stat.$\pm$ syst. \\
  \hline
  \input{a1.helium.tex}
  \end{tabular}

  \bigskip
  \begin{tabular}{|c|c|r|} 
  $\langle x \rangle$ & 
  $\langle Q^2\rangle\quad \left[({\rm GeV}/c)^2\right]$ & 
  $A_1^{h^+} \pm$ stat.$\pm$ syst. \\
  \hline
  \input{a1h+.helium.tex}
  \end{tabular}

  \bigskip
  \begin{tabular}{|c|c|r|}  
  $\langle x \rangle$ & 
  $\langle Q^2\rangle\quad \left[({\rm GeV}/c)^2\right]$ & 
  $A_1^{h^-} \pm$ stat.$\pm$ syst. \\
  \hline
  \input{a1h-.helium.tex}
  \end{tabular}
\end{table}

%
%

\begin{table}
 \centering
 \caption{The correlation coefficients $\rho$ between the asymmetries. 
  The proton and $^3$He asymmetries are uncorrelated. The correlation 
  coefficients between the inclusive and semi-inclusive asymmetries on 
  each target are given by  $\rho(A_1^{h^+},A_1^{h^-})=
  \langle n^+ n^- \rangle/  
  \sqrt{ \langle n^{+\ 2} \rangle \langle n^{-\ 2} \rangle}$ 
  and 
  $\rho(A_1,A_1^{h^{+(-)}})=
  \langle n^{+(-)}\rangle/
  \sqrt{\langle n^{+(-)\ 2}\rangle}$, where $n^{+(-)}$ is the number of 
  positively (negatively) charged hadrons per scattered positron.  
  \label{tab:asymmetry_correlations}}
  \bigskip
  \begin{tabular}{|c||cccccc|}
  $\langle x \rangle$ & $\rho(A_{\rm 1p   },A_{\rm 1p }^{h^+})$  
  & $\rho(A_{\rm 1p       },A_{\rm 1p }^{h^-})$  
  & $\rho(A_{\rm 1p }^{h^+},A_{\rm 1p }^{h^-})$  
  & $\rho(A_{\rm 1He      },A_{\rm 1He}^{h^+})$  
  & $\rho(A_{\rm 1He      },A_{\rm 1He}^{h^-})$  
  & $\rho(A_{\rm 1He}^{h^+},A_{\rm 1He}^{h^-})$\\
  \hline        
  \input{correlations.a1.tex}  
  \end{tabular}
\end{table}

%
%

\begin{table}
  \centering
  \caption{ The {\bf flavor decomposition} 
  $(\Delta u+\Delta \bar{u})/(u+\bar{u})$, 
  $(\Delta d+\Delta \bar{d})/(d+\bar{d})$, 
  and ${\Delta q_{\rm s}}/{q_{\rm s}}$ of the quark polarization 
  as a function
  of $x$ derived from the HERMES inclusive and semi-inclusive
  asymmetries on the $^3$He and proton targets. The first error is 
  statistical and the second one systematic. The values were obtained 
  with the assumption that the sea quark polarization is flavor symmetric.}
  \bigskip
  \begin{tabular}{|c||r|r|r|}
    $\langle x \rangle $ 
  & $ \frac{\Delta u + \Delta \bar u}{u + \bar u}\pm$ stat.$\pm$ syst. 
  & $ \frac{\Delta d + \Delta \bar d}{d + \bar d} \pm$ stat.$\pm$ syst. 
  & $ \Delta q_{\rm s}/q_{\rm s} \pm$ stat.$\pm$ syst. \\
  \hline
  \input{dq.fs.tex}
  \end{tabular}
\end{table}

\vspace*{0.5cm}
\begin{table}
  \centering
  \caption{Statistical correlation coefficients $\rho$  
           between the quark polarizations  
           $(\Delta u+\Delta \bar{u})/(u+\bar{u})$, 
           $(\Delta d+\Delta \bar{d})/(d+\bar{d})$, 
           and ${\Delta q_{\rm s}}/{q_{\rm s}}$ in each $x$-bin. 
           Also shown is the $\chi^2_{\rm min}$ of the fit. The number 
           of degrees of freedom of the fit is six for the bins 1 to 7 
           and seven for the bins 8 and 9. 
           \label{tab:deltaq_correl.fs}}
  \bigskip
  \begin{tabular}{|c||r|r|r||r|}
  $\langle x \rangle$ & 
  $ \rho\left(\frac{\Delta u + \Delta \bar u}{u + \bar u},
          \frac{\Delta d + \Delta \bar d}{d + \bar d}\right)$           
      & $ \rho\left(\frac{\Delta u + \Delta \bar u}{u + \bar u},
          \frac{\Delta q_{\rm s}}{q_{\rm s}}\right)$
      & $ \rho\left(\frac{\Delta d + \Delta \bar d}{d + \bar d},
          \frac{\Delta q_{\rm s}}{q_{\rm s}}\right)$
      & $ \chi^2_{\rm min}$\\
  \hline
  \input{correlations.dq.fs.tex}
  \end{tabular}
\end{table}
\clearpage

%
%

\begin{table}
  \centering
  \caption{The {\bf valence decomposition} $\Delta u_{\rm v}/u_{\rm v},
  \Delta d_{\rm v}/d_{\rm v}$ and $\Delta q_{\rm s}/q_{\rm s}$ 
  of the quark polarization as a function
  of $x$, derived from the HERMES inclusive and semi-inclusive
  asymmetries on the $^3$He and proton targets. The first error is 
  statistical and the second one systematic. The values were obtained 
  with the assumption that the sea quark polarization is flavor symmetric.    
  \label{tab:deltaq_all.val}}
  \bigskip
  \begin{tabular}{|c||r|r|r|}
  $\langle x \rangle$ & $ \Delta u_{\rm v}/u_{\rm v}     \pm$ stat.$\pm$ syst. 
      & $ \Delta d_{\rm v}/d_{\rm v}     \pm$ stat.$\pm$ syst. 
      & $ \Delta q_{\rm s}/q_{\rm s} \pm$ stat.$\pm$ syst. \\
  \hline
  \input{dq.val.tex}
  \end{tabular}
\end{table}

\vspace*{0.5cm}
\begin{table}
  \centering
  \caption{Statistical correlation coefficients $\rho$ between the quark 
  polarizations $\Delta u_{\rm v}/u_{\rm v},
  \Delta d_{\rm v}/d_{\rm v}$ and $\Delta q_{\rm s}/q_{\rm s}$ 
  in each $x$-bin. Also shown is the $\chi^2_{\rm min}$ of the fit. The 
  number of degrees of freedom of the fit is six for the bins 1 to 7 and 
  seven for the bins 8 and 9. 
  \label{tab:deltaq_correl.val}}
  \bigskip
  \begin{tabular}{|c||r|r|r||r|}
  $\langle x \rangle$ & $ \rho\left(\Delta u_{\rm v}/u_{\rm v},
                \Delta d_{\rm v}/d_{\rm v}\right)$           
      & $ \rho\left(\Delta u_{\rm v}/u_{\rm v},
                \Delta q_{\rm s}/q_{\rm s}\right)$
      & $ \rho\left(\Delta d_{\rm v}/d_{\rm v},
                \Delta q_{\rm s}/q_{\rm s}\right)$
      & $ \chi^2_{\rm min}$\\
  \hline
  \input{correlations.dq.val.tex}
  \end{tabular}
\end{table}

\end{document}

%% file: a1.proton.tex
%
%
%
%
%
%
$ 0.033 $ & $  1.21 $ & $   0.078 \pm    0.006 \pm    0.006 $ \\ 
$ 0.047 $ & $  1.47 $ & $   0.104 \pm    0.007 \pm    0.008 $ \\ 
$ 0.065 $ & $  1.72 $ & $   0.117 \pm    0.008 \pm    0.009 $ \\ 
$ 0.087 $ & $  1.99 $ & $   0.163 \pm    0.009 \pm    0.012 $ \\ 
$ 0.119 $ & $  2.30 $ & $   0.194 \pm    0.009 \pm    0.015 $ \\ 
$ 0.168 $ & $  2.66 $ & $   0.240 \pm    0.011 \pm    0.019 $ \\ 
$ 0.245 $ & $  3.06 $ & $   0.307 \pm    0.014 \pm    0.025 $ \\ 
$ 0.342 $ & $  3.74 $ & $   0.415 \pm    0.023 \pm    0.034 $ \\ 
$ 0.465 $ & $  5.16 $ & $   0.552 \pm    0.032 \pm    0.048 $ \\ 

%% file: a1h+.proton.tex
%
%
%
%
%
%
$ 0.033 $ & $  1.21 $ & $   0.087 \pm    0.016 \pm    0.006 $ \\ 
$ 0.047 $ & $  1.46 $ & $   0.111 \pm    0.017 \pm    0.008 $ \\ 
$ 0.065 $ & $  1.75 $ & $   0.136 \pm    0.017 \pm    0.010 $ \\ 
$ 0.087 $ & $  2.14 $ & $   0.180 \pm    0.019 \pm    0.013 $ \\ 
$ 0.118 $ & $  2.70 $ & $   0.247 \pm    0.020 \pm    0.018 $ \\ 
$ 0.165 $ & $  3.67 $ & $   0.245 \pm    0.024 \pm    0.019 $ \\ 
$ 0.238 $ & $  5.16 $ & $   0.411 \pm    0.033 \pm    0.031 $ \\ 
$ 0.339 $ & $  7.23 $ & $   0.403 \pm    0.066 \pm    0.033 $ \\ 
$ 0.447 $ & $  9.75 $ & $   0.619 \pm    0.132 \pm    0.053 $ \\ 

%% file: a1h-.proton.tex
%
%
%
%
%
%
$ 0.033 $ & $  1.21 $ & $   0.049 \pm    0.020 \pm    0.004 $ \\ 
$ 0.047 $ & $  1.46 $ & $   0.094 \pm    0.021 \pm    0.007 $ \\ 
$ 0.065 $ & $  1.75 $ & $   0.066 \pm    0.022 \pm    0.005 $ \\ 
$ 0.087 $ & $  2.14 $ & $   0.060 \pm    0.025 \pm    0.006 $ \\ 
$ 0.118 $ & $  2.70 $ & $   0.181 \pm    0.027 \pm    0.014 $ \\ 
$ 0.165 $ & $  3.67 $ & $   0.194 \pm    0.034 \pm    0.016 $ \\ 
$ 0.238 $ & $  5.16 $ & $   0.213 \pm    0.048 \pm    0.019 $ \\ 
$ 0.339 $ & $  7.23 $ & $   0.572 \pm    0.100 \pm    0.043 $ \\ 
$ 0.447 $ & $  9.75 $ & $   0.111 \pm    0.203 \pm    0.031 $ \\ 

%% file: a1.helium.tex
%
%
%
%
%
%
$ 0.033 $ & $  1.21 $ & $  -0.036 \pm    0.013 \pm    0.005 $ \\ 
$ 0.047 $ & $  1.47 $ & $  -0.009 \pm    0.014 \pm    0.003 $ \\ 
$ 0.065 $ & $  1.72 $ & $  -0.027 \pm    0.015 \pm    0.003 $ \\ 
$ 0.087 $ & $  1.99 $ & $  -0.025 \pm    0.018 \pm    0.003 $ \\ 
$ 0.119 $ & $  2.30 $ & $  -0.033 \pm    0.019 \pm    0.004 $ \\ 
$ 0.168 $ & $  2.66 $ & $  -0.037 \pm    0.023 \pm    0.005 $ \\ 
$ 0.245 $ & $  3.06 $ & $  -0.006 \pm    0.028 \pm    0.006 $ \\ 
$ 0.342 $ & $  3.74 $ & $   0.072 \pm    0.047 \pm    0.012 $ \\ 
$ 0.465 $ & $  5.16 $ & $  -0.021 \pm    0.066 \pm    0.019 $ \\ 

%% file: a1h+.helium.tex
%
%
%
%
%
%
$ 0.033 $ & $  1.21 $ & $  -0.051 \pm    0.032 \pm    0.007 $ \\ 
$ 0.047 $ & $  1.46 $ & $  -0.011 \pm    0.033 \pm    0.006 $ \\ 
$ 0.065 $ & $  1.75 $ & $  -0.030 \pm    0.034 \pm    0.006 $ \\ 
$ 0.087 $ & $  2.14 $ & $  -0.035 \pm    0.039 \pm    0.006 $ \\ 
$ 0.118 $ & $  2.70 $ & $  -0.024 \pm    0.041 \pm    0.006 $ \\ 
$ 0.165 $ & $  3.67 $ & $   0.006 \pm    0.048 \pm    0.006 $ \\ 
$ 0.238 $ & $  5.16 $ & $  -0.151 \pm    0.065 \pm    0.013 $ \\ 
$ 0.339 $ & $  7.23 $ & $  -0.088 \pm    0.133 \pm    0.013 $ \\ 
$ 0.447 $ & $  9.75 $ & $  -0.105 \pm    0.275 \pm    0.021 $ \\ 

%% file: a1h-.helium.tex
%
%
%
%
%
%
$ 0.033 $ & $  1.21 $ & $  -0.079 \pm    0.037 \pm    0.008 $ \\ 
$ 0.047 $ & $  1.46 $ & $   0.021 \pm    0.039 \pm    0.006 $ \\ 
$ 0.065 $ & $  1.75 $ & $  -0.021 \pm    0.042 \pm    0.006 $ \\ 
$ 0.087 $ & $  2.14 $ & $   0.027 \pm    0.049 \pm    0.006 $ \\ 
$ 0.118 $ & $  2.70 $ & $   0.009 \pm    0.052 \pm    0.006 $ \\ 
$ 0.165 $ & $  3.67 $ & $  -0.022 \pm    0.065 \pm    0.006 $ \\ 
$ 0.238 $ & $  5.16 $ & $   0.062 \pm    0.089 \pm    0.009 $ \\ 
$ 0.339 $ & $  7.23 $ & $  -0.220 \pm    0.191 \pm    0.019 $ \\ 
$ 0.447 $ & $  9.75 $ & $   0.293 \pm    0.616 \pm    0.028 $ \\ 

%% file: correlations.a1.tex
%
%
%
%
%
%
%
%
%
 $0.033 $ &   $0.452 $ &  $0.394 $ &   $0.130 $  &         $0.446 $ &   $0.395 $ &     $0.128 $\\ 
 $0.047 $ &   $0.490 $ &  $0.414 $ &   $0.140 $  &         $0.491 $ &   $0.417 $ &     $0.137 $\\ 
 $0.065 $ &   $0.517 $ &  $0.406 $ &   $0.134 $  &         $0.507 $ &   $0.411 $ &     $0.131 $\\ 
 $0.087 $ &   $0.509 $ &  $0.379 $ &   $0.120 $  &         $0.497 $ &   $0.386 $ &     $0.117 $\\ 
 $0.119 $ &   $0.464 $ &  $0.328 $ &   $0.108 $  &         $0.452 $ &   $0.336 $ &     $0.105 $\\ 
 $0.168 $ &   $0.375 $ &  $0.253 $ &   $0.098 $  &         $0.365 $ &   $0.260 $ &     $0.096 $\\ 
 $0.245 $ &   $0.267 $ &  $0.171 $ &   $0.084 $  &         $0.260 $ &   $0.178 $ &     $0.083 $\\ 
 $0.342 $ &   $0.188 $ &  $0.115 $ &   $0.066 $  &         $0.183 $ &   $0.120 $ &     $0.066 $\\ 
 $0.465 $ &   $0.130 $ &  $0.076 $ &   $0.050 $  &         $0.127 $ &   $0.080 $ &     $0.051 $\\

%% file: dq.fs.tex
%
%
%
%
%
%
%
 $  0.033$ & $  0.103 \pm   0.013 \pm   0.010$ & $ -0.126 \pm   0.052 \pm   0.021$ & $  0.012 \pm   0.095 \pm   0.042$  \\ 
 $  0.047$ & $  0.108 \pm   0.014 \pm   0.007$ & $ -0.063 \pm   0.057 \pm   0.016$ & $  0.109 \pm   0.103 \pm   0.039$  \\ 
 $  0.065$ & $  0.143 \pm   0.014 \pm   0.011$ & $ -0.127 \pm   0.063 \pm   0.017$ & $ -0.055 \pm   0.121 \pm   0.068$  \\ 
 $  0.087$ & $  0.192 \pm   0.016 \pm   0.013$ & $ -0.143 \pm   0.077 \pm   0.016$ & $ -0.215 \pm   0.169 \pm   0.117$  \\ 
 $  0.119$ & $  0.215 \pm   0.017 \pm   0.014$ & $ -0.209 \pm   0.090 \pm   0.026$ & $  0.148 \pm   0.236 \pm   0.119$  \\ 
 $  0.168$ & $  0.260 \pm   0.021 \pm   0.019$ & $ -0.239 \pm   0.123 \pm   0.050$ & $  0.200 \pm   0.441 \pm   0.137$  \\ 
 $  0.245$ & $  0.348 \pm   0.027 \pm   0.022$ & $ -0.235 \pm   0.191 \pm   0.078$ & $ -1.108 \pm   1.077 \pm   0.691$  \\ 
 $  0.342$ & $  0.374 \pm   0.043 \pm   0.025$ & $  0.478 \pm   0.388 \pm   0.085$ &                                    \\ 
 $  0.465$ & $  0.656 \pm   0.063 \pm   0.039$ & $ -0.989 \pm   0.787 \pm   0.129$ &                                    \\ 

%% file: correlations.dq.fs.tex
%
%
%
%

%
%
 $   0.033 $ & $  -0.76 $ & $  -0.12 $ & $  -0.38 $  & $    5.4 $ \\ 
 $   0.047 $ & $  -0.76 $ & $  -0.04 $ & $  -0.40 $  & $    1.0 $ \\ 
 $   0.065 $ & $  -0.76 $ & $  -0.02 $ & $  -0.42 $  & $    0.5 $ \\ 
 $   0.087 $ & $  -0.75 $ & $  -0.02 $ & $  -0.44 $  & $   11.2 $ \\ 
 $   0.119 $ & $  -0.73 $ & $  -0.05 $ & $  -0.46 $  & $    7.9 $ \\ 
 $   0.168 $ & $  -0.69 $ & $  -0.11 $ & $  -0.47 $  & $    5.1 $ \\ 
 $   0.245 $ & $  -0.57 $ & $  -0.31 $ & $  -0.45 $  & $   13.3 $ \\ 
 $   0.342 $ & $  -0.84 $ & $        $ & $        $  & $   11.6 $ \\ 
 $   0.465 $ & $  -0.85 $ & $        $ & $        $  & $    5.4 $ \\ 

%% file: dq.val.tex
%
%
%
%
%
%
%
 $  0.033$ & $  0.215 \pm   0.122 \pm   0.060$ & $ -0.421 \pm   0.306 \pm   0.113$ & $  0.012 \pm   0.095 \pm   0.039$  \\ 
 $  0.047$ & $  0.108 \pm   0.094 \pm   0.039$ & $ -0.354 \pm   0.274 \pm   0.111$ & $  0.109 \pm   0.103 \pm   0.040$  \\ 
 $  0.065$ & $  0.269 \pm   0.081 \pm   0.055$ & $ -0.225 \pm   0.266 \pm   0.106$ & $ -0.055 \pm   0.121 \pm   0.068$  \\ 
 $  0.087$ & $  0.374 \pm   0.080 \pm   0.062$ & $ -0.064 \pm   0.297 \pm   0.151$ & $ -0.215 \pm   0.169 \pm   0.119$  \\ 
 $  0.119$ & $  0.234 \pm   0.073 \pm   0.050$ & $ -0.519 \pm   0.320 \pm   0.098$ & $  0.148 \pm   0.236 \pm   0.122$  \\ 
 $  0.168$ & $  0.270 \pm   0.079 \pm   0.041$ & $ -0.520 \pm   0.418 \pm   0.114$ & $  0.200 \pm   0.441 \pm   0.148$  \\ 
 $  0.245$ & $  0.456 \pm   0.093 \pm   0.071$ & $  0.136 \pm   0.630 \pm   0.437$ & $ -1.108 \pm   1.077 \pm   0.717$  \\ 
 $  0.342$ & $  0.385 \pm   0.045 \pm   0.026$ & $  0.602 \pm   0.489 \pm   0.264$ &                                    \\ 
 $  0.465$ & $  0.663 \pm   0.064 \pm   0.040$ & $ -1.127 \pm   0.897 \pm   0.220$ &                                    \\ 

%% file: correlations.dq.val.tex
%
%
%
%
%
%
 $   0.033 $ & $   0.74 $ & $  -0.98 $ & $  -0.87 $  & $    5.4 $ \\ 
 $   0.047 $ & $   0.71 $ & $  -0.96 $ & $  -0.86 $  & $    1.0 $ \\ 
 $   0.065 $ & $   0.70 $ & $  -0.96 $ & $  -0.86 $  & $    0.5 $ \\ 
 $   0.087 $ & $   0.71 $ & $  -0.96 $ & $  -0.87 $  & $   11.2 $ \\ 
 $   0.119 $ & $   0.72 $ & $  -0.95 $ & $  -0.88 $  & $    7.9 $ \\ 
 $   0.168 $ & $   0.75 $ & $  -0.95 $ & $  -0.90 $  & $    5.1 $ \\ 
 $   0.245 $ & $   0.79 $ & $  -0.96 $ & $  -0.92 $  & $   13.3 $ \\ 
 $   0.342 $ & $  -0.84 $ & $        $ & $        $  & $   11.6 $ \\ 
 $   0.465 $ & $  -0.85 $ & $        $ & $        $  & $    5.4 $ \\ 